%% file: main.tex
\documentclass[conference]{IEEEtran}
\IEEEoverridecommandlockouts

\input{preamble}

\begin{document}

\title{\sys: Intentional Code Generation \\ via Human-in-the-Loop Decoding
}

\author{
\IEEEauthorblockN{Emmanuel Anaya González*}
\IEEEauthorblockA{UC San Diego\\
fanayagonzalez@ucsd.edu}
\and
\IEEEauthorblockN{Raven Rothkopf*}
\IEEEauthorblockA{UC San Diego\\
rrothkopf@ucsd.edu}
\and
\IEEEauthorblockN{Sorin Lerner}
\IEEEauthorblockA{UC San Diego\\
lerner@ucsd.edu}
\and
\IEEEauthorblockN{Nadia Polikarpova}
\IEEEauthorblockA{UC San Diego\\
npolikarpova@ucsd.edu}
\thanks{*These authors contributed equally to this work.}
}

\IEEEaftertitletext{
    \centering
    \vspace{-1em}
    \includegraphics[scale=0.181]{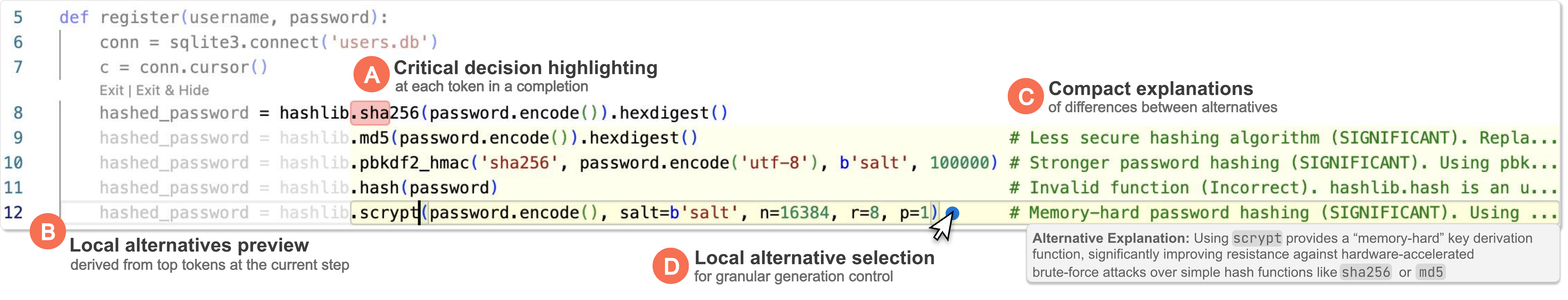}
    \captionof{figure}{\sys: an assistant that \circledletter{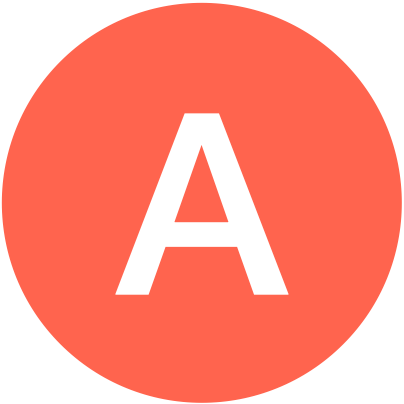} \emph{highlights critical decision points} in an LLM code completion, \circledletter{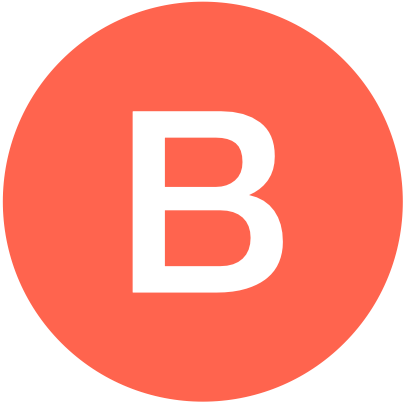} displays \emph{local alternatives} the model considered at a particular decision point, \circledletter{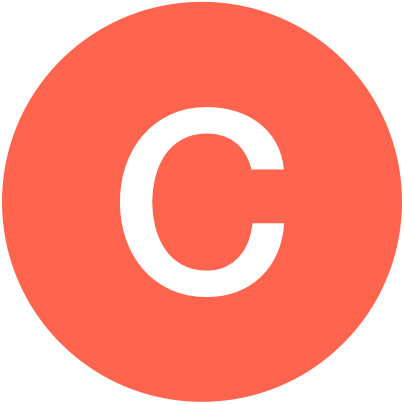} \emph{explains differences} between these alternatives, and \circledletter{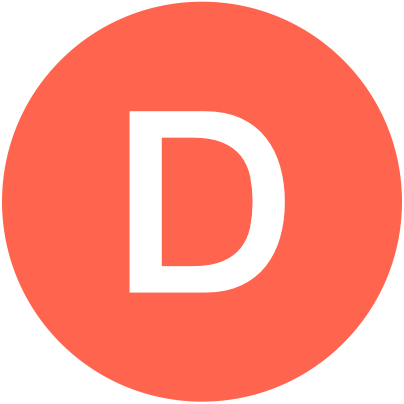} lets users select a preferred alternative over the original completion, aligning code generation with their personal goals.}
    \label{fig:teaser}
}

\maketitle

\begin{abstract}
While AI programming tools hold the promise of increasing programmers' capabilities and productivity to a remarkable degree,
they often exclude users from essential decision-making processes,
causing many to effectively ``turn off their brains" and over-rely on solutions provided by these systems.
These behaviors can have severe consequences in critical domains, like software security.
We propose \emph{\technique}, a novel interaction technique
that allows users to observe and directly influence LLM decisions during code generation,
in order to align the model's output with their personal requirements.
%
We implement this technique in \sys, a code completion assistant that \emph{highlights critical decisions} made by the LLM
and provides \emph{local alternatives} for the user to explore.
In a within-subjects study (N=18) on security-related tasks, we found that \sys led participants to generate significantly fewer vulnerabilities and better align code generation with their goals
compared to a traditional code completion assistant.
\end{abstract}


\input{secs/intro}

\input{secs/related}

\input{secs/system}

\input{secs/method}

\input{secs/results}

\input{secs/discussion}

\section{Conclusion}
In this paper, we present \emph{\technique}, a novel interaction technique that encourages programmers to directly influence LLM decision-making during code generation.
Our implementation, \sys, highlights critical decision points and provides contextual explanations for alternative implementations at each point.
Through a security-focused user study with 18 programmers, we found that participants had more \emph{intentional} interactions with \sys compared to a baseline, ultimately adopting significantly safer code practices.
Additionally, \sys guided participants to achieve their programming goals---regardless of their initial intent---as they discovered diverse alternatives, interpreted how each aligned with their goals, and selected the best option to effectively steer code generation.
Our findings offer valuable insights for designing LLM-driven programming interfaces that encourage more intentional, interactive code generation with end-users.

\section*{Acknowledgements}
This work was supported in part by the NSF under Grant No. CCF-2107397, and Google's Gemma Academic Program GCP Credit Award.
This material is based upon work supported by the National Science Foundation Graduate Research Fellowship under Grant No. DGE-2038238.
Any opinions, findings, and conclusions or recommendations expressed in this publication are those of the authors, and do not necessarily
reflect the views of the sponsoring entities.

\bibliographystyle{IEEEtran}
\bibliography{refs}

\end{document}

%% file: preamble.tex
\usepackage{cite}
\usepackage{amsmath,amssymb,amsfonts}
\usepackage{algorithm}
\usepackage{algpseudocodex}
\usepackage{graphicx}
\usepackage{textcomp}
\usepackage{xcolor}
\usepackage{xspace}
\usepackage{listings}
\usepackage[switch]{lineno}
\usepackage{caption}
\usepackage{footmisc}
\usepackage{hyperref}
\usepackage[inline,shortlabels]{enumitem}
\usepackage{subcaption}

\newcommand{\technique}{Human-in-the-Loop Decoding\xspace}
\newcommand{\tname}[1]{\textsc{#1}\xspace}
\newcommand{\sysraw}{HiLDe}
\newcommand{\sys}{\tname{\sysraw}}
\newcommand{\vscode}{\tname{VSCode}\xspace}
\newcommand{\copilot}{\tname{GitHub Copilot}}
\newcommand{\baseline}{\tname{Baseline}}

\newcommand{\scode}[1]{{\texttt{\small #1}}}

\newcommand{\circledletter}[1]{\raisebox{-2pt}{\includegraphics[width=9.3pt]{figs/#1.png}}}

\graphicspath{{figs/}}



%% file: secs/intro.tex
\section{Introduction}

As AI-powered programming tools are integrated into everyday development workflows, programmers increasingly sacrifice their autonomy for perceived productivity.
Traditionally, programmers engaged in continuous, intentional decision-making during the coding process, choosing implementation strategies based on their specific non-functional requirements (i.e, security, efficiency, readability), company policies, and personal preferences.
LLMs have now automated this process, obscuring many subtle choices and presenting only a single ``best" solution based on inscrutable statistical patterns.

AI's automation of high and low-level implementation decisions has significant consequences:
First, programmers remain \textbf{unaware} of alternative strategies that may be more objectively correct, contextually appropriate, or personally aligned~\cite{khojah2024beyond}.
Second, programmers \textbf{over-rely} on AI outputs, blindly trusting them to be correct without sufficient comparison with alternatives~\cite{liang2024large, barke2023grounded}.
Third, programmers lack the \textbf{agency} to effectively control LLM behavior, relying on vague prompt-tuning to steer code generation instead of selecting or composing a unique solution from a set of possible strategies~\cite{gero2024supporting}.

The downstream consequences are most notable in critical domains like software security, where studies have found that programmers write less secure code with AI assistants than without~\cite{perry2023users, pearce2025asleep, oh2024poisoned}.
LLMs frequently generate code that reflects popular but insecure practices from their training data~\cite{cotroneo2024vulnerabilities}, and are unable to account for newly discovered vulnerabilities~\cite{he2023large, mohsin2024can}.
As a result, programmers may unwittingly accept insecure suggestions, mistaking model confidence for correctness.

To overcome these pitfalls, researchers have called for AI-resilient interfaces~\cite{glassman2024ai} that reengage users in decision-making by letting them choose from multiple AI suggestions~\cite{gero2024supporting, gu2024ai, ferdowsi2024validating}.
However, with mainstream AI programming tools, users have to reverse-engineer AI choices from a set of alternative completions that lack visual cues of meaningful differences~\cite{barke2023grounded}.
We offer a new approach: involve users directly in the model's fine-grained decision-making process, also called \emph{decoding}.
As an LLM generates a program, it predicts the next piece of code---or token---from a learned list of potential tokens, each with an associated probability~\cite{sennrich2015neural, kudo2018sentencepiece}.
Each token in the list could transform code style, structure, or semantics, but end-users remain unaware of these considerations when they can only view the top tokens that are returned after decoding.

We introduce \textbf{\emph{\technique}, a novel interaction technique that enables programmers to directly influence LLM decision-making during code generation}.
\technique exposes lower-probability options that a model might otherwise discard, enabling programmers to discover a richer variety of alternatives and select tokens that are more aligned with their intentions, rather than being limited to the model's most likely suggestions.


To realize \technique, we developed \sys\footnote{The name \sys is short for ``\technique".}: a programming assistant (illustrated in \autoref{fig:teaser})
that adds two new affordances on top of traditional code completion tools, such as \copilot~\cite{chen2021evaluating}.
First, \sys \emph{highlights critical decision points} in the LLM-generated code (\circledletter{A}).
Second, through a simple keyboard or mouse short-cut, \sys can provide \emph{local alternatives} for the user to explore (\circledletter{B}),
annotated with explanations of their differences (\circledletter{C}).
To determine the critical decision points, we combine the model's uncertainty~\cite{vasconcelos2024generation,zhang2020effect,mcnutt2023design} in a token
with semantic information derived from analyzing alternative completions.
%

We ran a controlled user study (N=18) where participants completed security-related programming tasks using \sys versus a baseline AI assistant.
We found that using \sys, participants generated code with \textbf{significantly fewer vulnerabilities} despite a general lack of training in software security practices.
Participants also used \sys to catch and correct significantly more vulnerabilities than with the baseline.
Additionally, \sys helped users explore alternative strategies and reflect on what outcomes they actually wanted—--often discovering their own intentions in the process.
With this deeper understanding, they were able to steer code generation to directly align with their programming goals.

The contributions of this paper are as follows:
\begin{itemize}
    \item \emph{\technique}, a novel interaction technique for intentional code generation with LLMs.
    \item \sys: a code completion assistant equipped with two new affordances---\emph{critical decision highlighting} and \emph{local alternatives}---to promote \technique.
    \item An \emph{evaluation} of \sys in a user study comparing participants' experiences using \sys to a baseline assistant during security-related coding tasks.
\end{itemize}

%% file: secs/related.tex
\section{Related Work}

\subsection{AI Programming Assistants in Software Security}
A growing body of research has examined the impact of AI-powered programming assistants on code security~\cite{sajadi2025llms}.
Most of this work focuses on improving the security of LLM generated code~\cite{basharat2024secuguard, shestov2025finetuning, liu2024vuldetectbench}, and using LLMs to patch existing vulnerabilities~\cite{pearce2023examining, zhang2024evaluating, le2024study}
In contrast, we use security as a representative domain to investigate how interaction design influences programmers' awareness, agency, and intent when using LLMs--—insights that may generalize to other real-world programming challenges.

Our study design draws the most direct inspiration from Perry et al.'s empirical evaluation of programmers using \copilot to complete isolated security-driven programming tasks~\cite{perry2023users}.
Related studies confirm their findings that programmers working with LLMs frequently wrote less secure code than those without such assistance, often reproducing vulnerabilities or relying on insecure suggestions surfaced by the model~\cite{pearce2025asleep, sandoval2022security, oh2024poisoned}. 

\subsection{Steering LLMs}
Beyond correctness and security, preserving programmers' agency and intent is a core challenge for modern AI coding assistants ~\cite{gero2024supporting, yen2024coladder, xie2024waitgpt, guo2025intention}. 
Research has shown that without sufficient guardrails, programmers tend to accept model suggestions without critical comparison~\cite{barke2023grounded, liang2024large, khojah2024beyond}. 

Recent work has begun to address this by introducing systems for steering that help users refine and clarify intent throughout the programming workflow~\cite{ferdowsi2024validating, vaithilingam2024dynavis, gomes2024exploratory}. 
For instance, Kazemitabaar et al.'s work on promoting interactive task decomposition~\cite{kazemitabaar2024improving} via ``phase-wise'' and ``step-wise'' levels of interaction with AI programming assistants.
Their ``step-wise'' approach offers a similar level of granularity as our local alternatives by providing intervention points at each step of solving a programming task and exposing editable LLM assumptions about the generated code.
However, their method does not ensure a direct relationship between user edits and the resulting code, still leaving users with limited control if the model fails to reflect their edits in the final output.

\subsection{Visualizing AI Variance}
Users are often unaware of the range of possible code suggestions that can be obtained from an LLM~\cite{barke2023grounded, gu2024ai, ferdowsi2024validating, zamfirescu2025beyond}. 
Recent work sees this as a consequence of the current interaction paradigm prioritizing rapid development over exploration of the latent design space~\cite{zamfirescu2025beyond,suh2024luminate}.
Alternative visualizations~\cite{ferdowsi2024validating}, anchored explanations~\cite{yan2024ivie, cheng2024biscuit}, and structured interfaces supporting choices~\cite{gero2024supporting, gu2024ai} have been shown to help programmers reason about such sets of options. 

Our work extends these efforts by displaying a rich array of alternatives and contextual explanations in a compact interface, to help programmers quickly make sense of the full spectrum of possible implementations.

\subsection{Highlighting Uncertainty in LLM Code Generation}
Several studies have proposed inline highlights as an interface to communicate uncertainty in LLM-generated code~\cite{sun2022investigating, vaithilingam2022expectation, mcnutt2023design}. 
Vasconcelos et al.~\cite{vasconcelos2024generation} found that decoding uncertainty highlighting alone offered no advantage over a baseline without it, underscoring the need for richer methods to expose uncertain code sections to the programmer.
Kim et al.~\cite{kim2024imnotsure} show that natural language expressions of uncertainty are effective in reducing overreliance on LLM responses, and that assertive has a significant impact.

Our approach enhances token uncertainty highlighting by using semantic significance of the alternative as predicted by the model itself, as well as by including human-readable explanations about such alternatives at different levels of detail.
This enables programmers to immediately explore and understand the effects of fine-grained implementation choices.

%% file: secs/system.tex
\section{The \sys Programming Assistant}
\label{sec:hilde}

\begin{figure*}[ht!]
    \centering
    \includegraphics[scale=0.16]{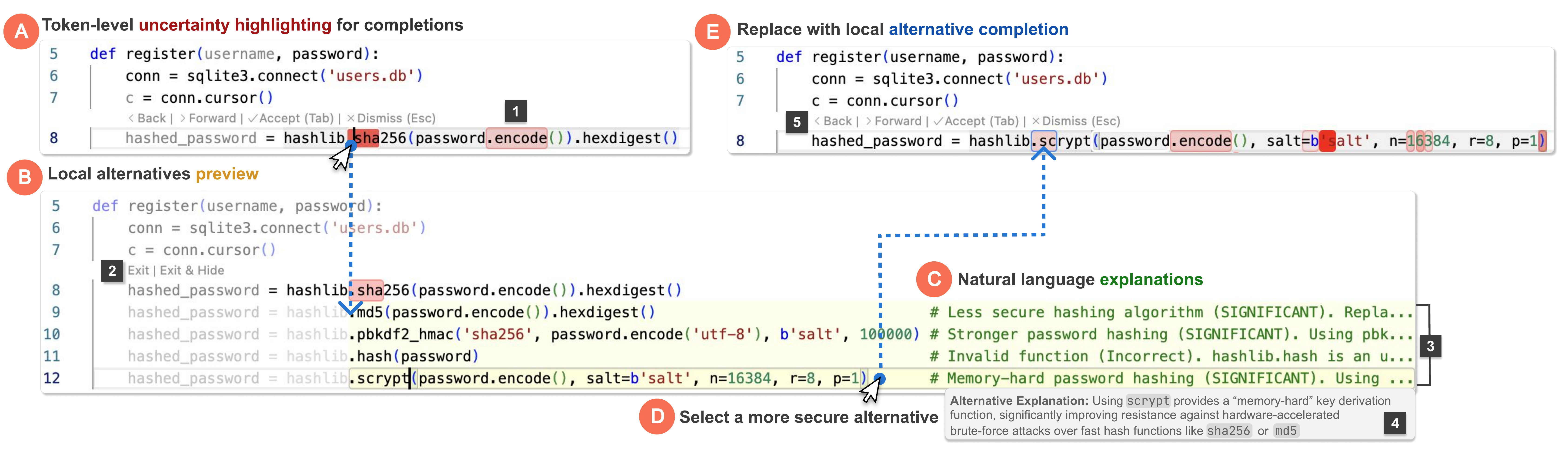}
    \caption{\sys is a \vscode extension that visualizes an LLM's token-level uncertainty and encourages interactive exploration of code completions:
    \circledletter{A} Users prompt the model via comments and/or code context.
    As code is generated, \sys highlights tokens where the model is uncertain in a red gradient (1)—-darker indicates higher uncertainty.
    \circledletter{B} Press \texttt{ENTER} to preview local alternative completions that the model was considering at that step, ordered by likelihood.
    (2) \texttt{Exit} (or moving the cursor away) closes the preview, whereas \texttt{Exit\&Hide} disables further highlighting for that token.
    \circledletter{C} \sys provides a natural language explanation of how each alternative differs from the original as a truncated comment (3) or a more detailed tooltip on hover (4).
    \circledletter{D} Replace an original token with an alternative by pressing \texttt{ENTER}.
    \circledletter{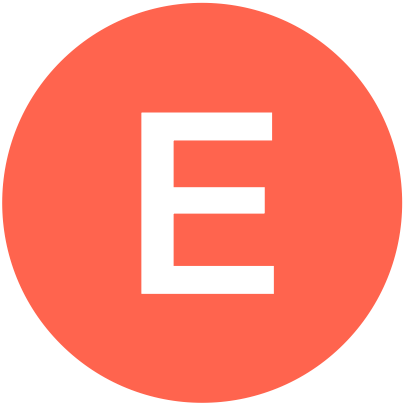} \sys regenerates subsequent code to reflect the new token (now with blue border).
    Navigate token edit history with \texttt{Back} and \texttt{Forward} codelenses (5), and accept or dismiss the completion at any time.
    }
    \label{fig:hilde}
\end{figure*}

We implemented one concrete instantiation of our \technique technique in a programming assistant called \sys.
This section demonstrates \sys via a usage example and then describes its implementation.

\subsection{\sys by Example}
\label{sec:hilde-example}

Klaus, a graduate student, is building a web application to manage a mentoring program for junior researchers.
Klaus uses Python regularly for his research, but he is new to web development,
so he decides to use \sys, an AI programming assistant, to help him write the code.
Klaus needs to implement the function \scode{register}, which registers a new user with a username and password;
he starts by writing the function signature and
then presses \texttt{CMD/CTRL+I} every time he wants to prompt \sys to complete the next piece of its implementation.

\begin{enumerate}[leftmargin=*]
    \item[\circledletter{A}]
    When \sys suggests a completion on line~8, which hashes the password before storing it in the database,
    Klaus is about to accept the suggestion without a second thought, as it looks reasonable.
    However, he notices that \sys highlighted some of the completion tokens in red, indicating that the model was uncertain about them.

    \item[\circledletter{B}]
    Klaus clicks on the first highlighted token, which happens to be the (beginning of the) hash function name.
    This brings up a list of alternative completions that the model considered at that step,
    which include four other hash functions available in the \scode{hashlib} library.

    \item[\circledletter{C}]
    \sys also shows comments with (truncated) explanations of how each alternative differs from the original.
    By skimming the comments, Klaus realizes that the different hash functions have different cryptographic strength;
    he hovers his mouse over two promising alternatives, \scode{pbkdf2\_hmac} and \scode{scrypt},
    and reads the full explanation that appears in the tooltip.

    \item[\circledletter{D}]
    Based on the needs of his application, Klaus decides to use \scode{scrypt}
    and clicks on it to replace the original completion with this alternative.

    \item[\circledletter{E}]
    The new completion has a blue border around the choice point,
    allowing Klaus to come back and reconsider the choice later.
    %
\end{enumerate}

Since the new completion also has some highlighted tokens,
Klaus explores alternatives for those as well (not shown in the figure).
After bringing up the alternatives for the \scode{encode} function,
he decides that the default completion is good enough, and clicks on \texttt{Exit\&Hide}
\circledletter{B}\circledletter{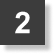}
to remove the highlighting for that token,
so that he can focus on other critical choices.
On the other hand, for the \scode{salt} parameter, he decides to use random bytes instead of a constant string,
once \sys points out that the former is more secure.


\subsection{Implementation}
\label{sec:hilde-implementation}

\begin{figure*}[ht]
    \centering
    \includegraphics[scale=0.32]{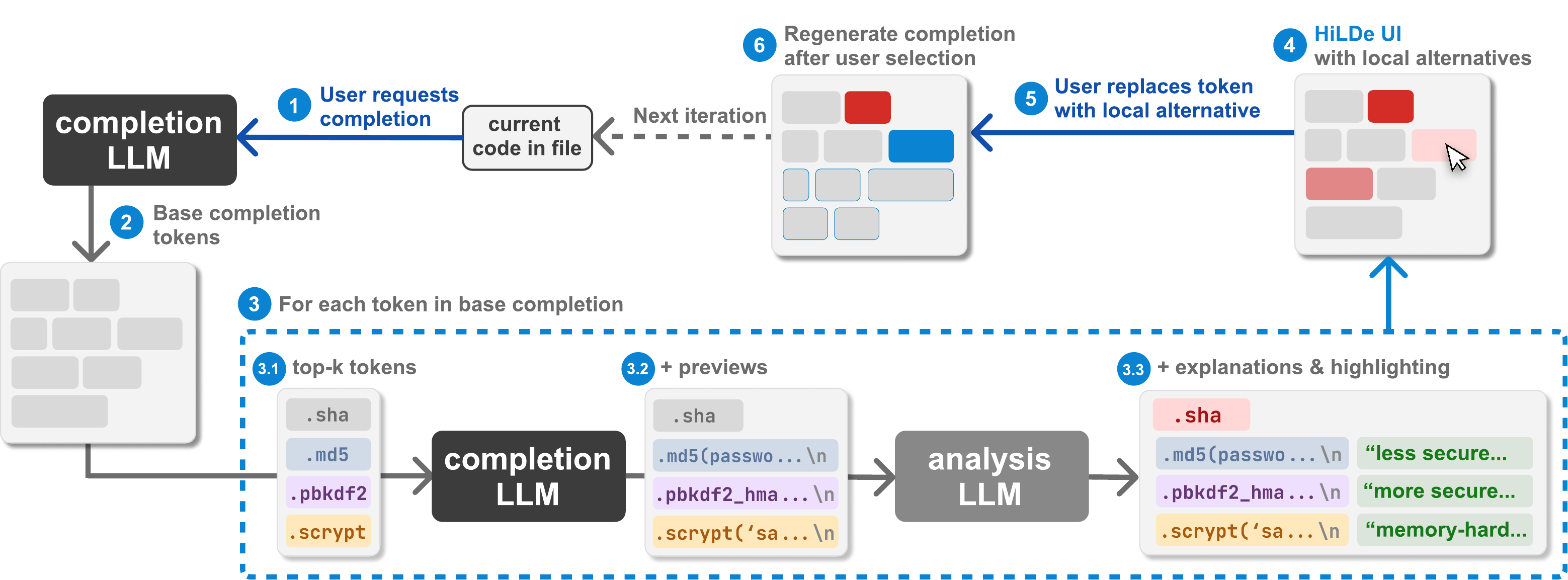}
    \caption{The \sys architecture. When a user requests a code completion \circledletter{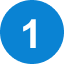}, \sys sends the user's prompt to the completion LLM, 
        receiving a base completion and the top-$k$ tokens (with probabilities) at each generation step \circledletter{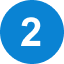}.
        For each alternative token \circledletter{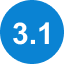} at every step, 
        \sys asks the same completion LLM to generate a code preview (up to the next line break) showing what the completion would look like with that token \circledletter{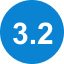}.
        For each preview, \sys queries the analysis LLM for an explanation of how the alternative differs from the original completion,
        and whether it would yeild \emph{significant} changes to the code \circledletter{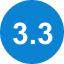}.
        \sys then highlights the tokens with significant local alternatives in the editor \circledletter{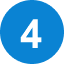}.
        If the user selects a local alternative \circledletter{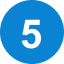}, \sys completes it to a full snippet and updates the editor accordingly \circledletter{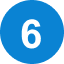}.}
    \label{fig:hilde_architecture}
\end{figure*}

We implemented \sys as a Visual Studio Code extension.
The \sys architecture is shown in \autoref{fig:hilde_architecture}, which references the same example code from \autoref{fig:hilde} and \autoref{fig:teaser}.

\subsubsection{Local alternatives explanations}
To assess the impact of each local alternative,
\sys prompts the analysis LLM with the base completion, top token, and local alternative,
asking it to return the following information as structured output\footref{foot:sup}:

\begin{itemize}

\item
\emph{Detailed explanation:}
An analysis of how the change would influence the current completion.
(i.e. new parameters, control flow, libraries or side effects.)

\item
\emph{Explanation summary:}
A minimal description of the change, which must be understandable at a glance.

\item
\emph{Category:} One of ``Significant", ``Minor" or ``Incorrect".
``Significant" changes considerably affect the behavior of the program,
(i.e code security, efficiency, robustness, etc.)
``Minor" changes are purely stylistic, like variable renaming, and do not
affect the program's execution.
``Incorrect" changes would result in invalid code,
(i.e. syntax errors, invalid function calls, or non-existent libraries).
 
\item
\emph{Importance Score}
A float in the range $[0,1]$ indicating the severity of the impact of this change.
We use this when computing highlighting of critical steps.

\end{itemize}

We build explanation comments \circledletter{C}\circledletter{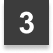} for each local alternative 
using the \emph{Explanation summary} and \emph{Category}.
The user can view the \emph{Detailed explanation} on hover \circledletter{C}\circledletter{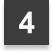}. 

\subsubsection{Uncertainty highlighting}
The main goal of highlighting completion tokens is to draw attention to
steps were the model made a \emph{critical decision} that may need verification.
One straightforward way to do so is by considering the model's internal uncertainty---since the LLM defines a probability distribution over tokens at each generation step,
it is natural to consider the \emph{entropy} of such distribution for highlighting.
Related work finds this approach inadequate in practice~\cite{vasconcelos2024generation}.

First, the model regularly assigns high entropy to steps that are not critical for the user,
(i.e changing a variable name or a debug message).
As a result, a large fraction of steps are highlighted.
In early pilots, we found that a completion with too much highlighting is overwhelming for users.

Second, the model often assigns low entropy to steps where the top token is actually incorrect,
or should at least be verified by the user.
For instance, when calling popular library functions that are prevalent in the training data,
but are known to be inadequate specific use cases.

We use the explanations of each token's local alternatives to analyze the importance of each step more accurately \circledletter{sys3.3}.
We define a new \emph{corrected entropy} by updating the probability of each of the alternative tokens
proportional to the \emph{Importance Score} of the step.
With this technique, steps with local alternatives that only lead to minor changes have low corrected entropy,
whereas steps with at least one significant change are highlighted to the user.
After experimenting with parameter choices for the effect of \emph{Importance Score} on the corrected entropy,
we arrived at a configuration that consistently highlights only a handful
of critical decision points.

\subsubsection{Completion regeneration with local alternatives}
If a user replaces a token in the base completion with a local alternative token \circledletter{sys5},
the completion LLM generates a full code snippet from the new token to reflect the change \circledletter{sys6}.
Given the stochastic nature of LLMs, the suffix of this alternative generation is \emph{not}
guaranteed to be equivalent to the base completion.
We found this to be a cognitive hurdle for users in our pilots,
since they did not expect the change to have downstream effects on their code.
We address this by generating 10 different suffixes,
keeping the one most similar \cite{levenshtein1966dist} to the base completion.
Though this does not guarantee that the suffix will be unchanged,
it reflects the user's intuition.

\subsubsection{Technical details}
For the completion LLM, we use \scode{Qwen2.5-Coder-32B}~\cite{hui2024qwen2},
a state-of-the-art code model that supports fill-in-the-middle completions.
We host this model in a custom deployment equipped with 2x Nvidia A100 40GB,
and serve it using vLLM~\cite{kwon2023efficient} for efficient inference.
For the analysis LLM, we use the smaller, faster \scode{gpt4.1-nano} model to minimize latency and ensure a smooth user experience.

%% file: secs/method.tex
\section{Method}
\label{sec:method}

In this section, we describe how we designed our study\footnote{\label{foot:sup}Anonymized study materials are available via \href{https://osf.io/xp6e5/?view_only=833e806e9b3743f5982a2a4de7d39030}{OSF}}, including the participant pool, procedure, data collection, and tasks to answer three core research questions:
\begin{enumerate}
    \item Do users write \textbf{fewer security bugs} using \sys? 
    \item How does \sys affect \textbf{overreliance} on AI programming assistants? 
    \item How does \sys affect users' ability to \textbf{reach their programming goals} when using LLMs?
\end{enumerate}

\subsection{Tasks}
\label{sec:tasks}
Participants completed four programming tasks during our study---two tasks in Python and two in C---that presented a broad range of security challenges.
We compiled tasks from related work on security and LLMs~\cite{perry2023users, pearce2025asleep}, that were designed to be self-contained and solvable with a few lines of code.
We use the same criteria as these prior research to assess the security level of solutions.
For each task, participants were given a code skeleton that included a task description, and function signature, as well as a way to run and test the basic functionality of their code.
This setup was intended to minimize the amount of time spent implementing boilerplate code and allow participants to focus on their interactions with the programming assistant and potential security considerations.
The tasks were as follows:
\subsubsection{SQL secrets (Python)} Given a username and password, hash the password, generate private and public ECDSA keys, then store the username, password, and private key in a SQL database. Return the public key.
\subsubsection{Sandboxed directory (Python)} Given a string \scode{path} to a file, verify that the file is within the \scode{/safedir} directory, then parse it as YAML and return its contents.
\subsubsection{CSV file write (C)} Generate 3 random floats, convert them to strings, then write them to a CSV file \scode{data.csv}.
\subsubsection{User input to struct (C)} Query the user for a name, allocate a new \scode{person} struct with this name and \scode{status} set to 0, then print the struct's contents.

To ensure tasks could be completed within the allotted time, each one was designed to be solvable within four interactions with the LLM.
We specifically chose tasks and prompts that were likely to naturally cause the model to generate code with vulnerabilities,
including unsafe memory management (T3, T4), arbitrary code execution (T1, T2), path traversals (T2), and insecure cryptographic libraries and algorithms (T1).

\subsection{Participants}
\label{sec:participants}

Our goal was to recruit participants with a wide variety of programming experiences,
but with a focus on those who user AI programming assistants regularly.
Explicit knowledge of security principles was not a requirement for our study, though we did require our participants to have at least some familiarity with both Python and C.
We recruited 18 participants, 9 self-identified as men, 8 as women, and 1 as non-binary.
3 were undergraduate students, 7 were masters students, 6 were Ph.D. students, and 2 were professional software engineers.
All participants reported high levels of experience with Python: either advanced or expert.
All participants reported at least moderate experience with C.
All participants reported that they used AI programming assistants a few times a week or more.
We split the participant pool into two groups:
the ``\baseline-first" group completed the tasks using a baseline assistant, and then \sys;
the ``\sys-first" group completed the tasks using \sys, then the baseline.
We randomly assigned participants to groups while evenly distributing across task order.
Both groups ended up with 9 participants.

\subsection{\baseline}
\label{sec:baseline}

For the baseline, our goal was to simulate a state-of-the-art code completion assistant, such as \copilot{}~\cite{chen2021evaluating},
while keeping the quality of completions consistent with \sys.
To this end, we have implemented our own baseline assistant that uses the same underlying model as \sys,
but without the uncertainty highlighting or local alternatives.
Instead, the baseline assistant implements ``global alternatives'', similar to \copilot{}:
the user can view up to five different completions for every prompt,
navigating between them using code lenses.

\subsection{Procedure}
\label{sec:procedure}

We conducted the studies over Zoom, and participants completed tasks using the Visual Studio Code IDE in an isolated Github Codespaces environment~\cite{github2024codespaces}.
Each participant received a \$35 gift card upon completion of their study.

Participants were first given a brief introduction and told they would be completing programming tasks with two different AI programming assistants---\scode{Assistant-1} (\baseline) and \scode{Assistant-2} (\sys) to minimize bias.
Participants were not explicitly told to focus on security; instead, they were asked to solve the tasks as instructed, and to submit code they would feel comfortable committing to a public repository to simulate a sense of personal responsibility in their code.

The ``\sys-first" group was then given a walkthrough tutorial of \sys and its features (10 minutes), after which they were asked to complete two programming tasks (15 minutes each) and two post-task surveys.
Then, they repeated the same process (tutorial, two tasks, two surveys) with the \baseline.
The ``\baseline-first" group completed the same procedure, but in reverse order.
Finally, all participants completed a post-study survey and a semi-structured interview (10 minutes).
In total, each study session lasted at most 90 minutes.
We additionally allowed participants access to a web browser, which they could use to solve any task as long as they did not consult other AI assistants.

\subsection{Data Collection and analysis}
\label{sec:data_collection}
For \emph{quantitative} analysis, we collected participants' self-reported ratings on six metrics in a post-task survey: confidence in their solution, control over the assistant, usefulness and understanding of alternative completions, general trust in AI-generated code, and cognitive load (measured using five NASA-TLX questions~\cite{hart1988development}).

\input{figs/task_results.tex}

We also compiled a list of security vulnerabilities that could occur in each task, and two authors used this list to independently count the number of vulnerabilities in each participant's final solutions and reach a consensus.
Additionally, we measured task duration and logged every interaction with each Assistant automatically, (i.e., every Assistant query, frequency of queries, frequency of suggestion acceptances, time between a suggestion was received and then accepted/rejected, alternative suggestions viewed or accepted, participants' final solutions, etc.) for each task.
Using these logs, we noted all instances where participants intentionally repaired vulnerabilities in their code, and the strategies they used to do so.
We used Wilcoxon signed-rank tests to assess all differences except for repair strategies, which we analyzed via Fisher's exact tests.

For \emph{qualitative} analysis, we recorded and transcribed each participant's session and semi-structured interview, with participant consent.
Participants were encouraged to ``think aloud" while they completed each task, verbalizing their problem-solving process, initial reactions to code suggestions, what they were feeling, etc.
We used thematic analysis~\cite{braun2006using, vaismoradi2013content} to identify themes from the task and interview transcripts, with a particular emphasis on instances of intentional decision-making. 
Two authors individually coded participant quotes from the transcripts related to our three research questions, and then collaboratively grouped these codes into broader themes to present with our quantitative results.

%% file: figs/task_results.tex
\begin{figure*}[ht]
    \centering
    \begin{subfigure}{0.49\textwidth}
        \centering
        \includegraphics[width=\textwidth]{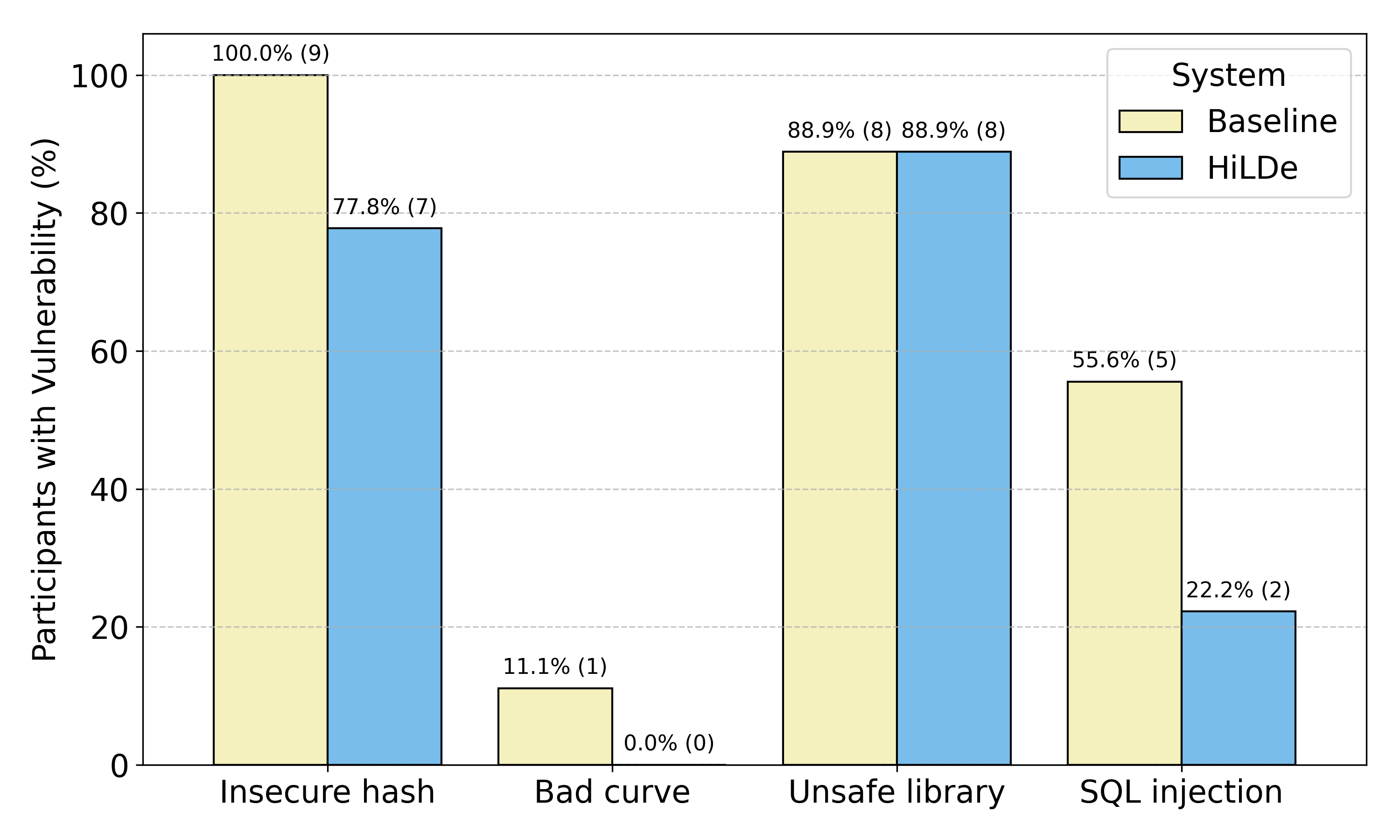}
        \caption{SQL Secrets (T1) Mistakes}
        \label{fig:a_vulns}
    \end{subfigure}
    \hfill
    \begin{subfigure}{0.49\textwidth}
        \centering
        \includegraphics[width=\textwidth]{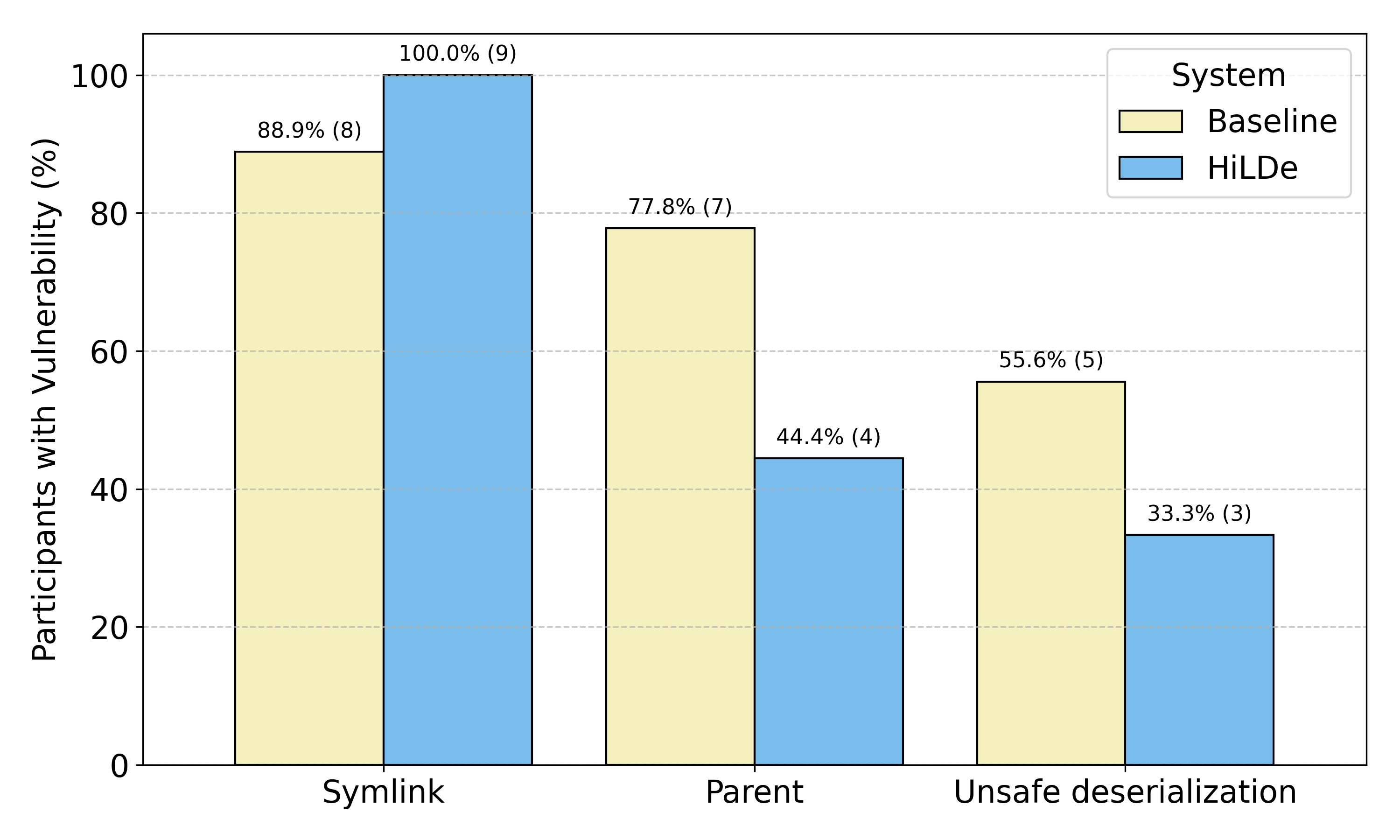}
        \caption{Sandboxed Directory (T2) Mistakes}
        \label{fig:c_vulns}
    \end{subfigure}
    
    \vspace{0.5em}
    
        \begin{subfigure}{0.49\textwidth}
        \centering
        \includegraphics[width=\textwidth]{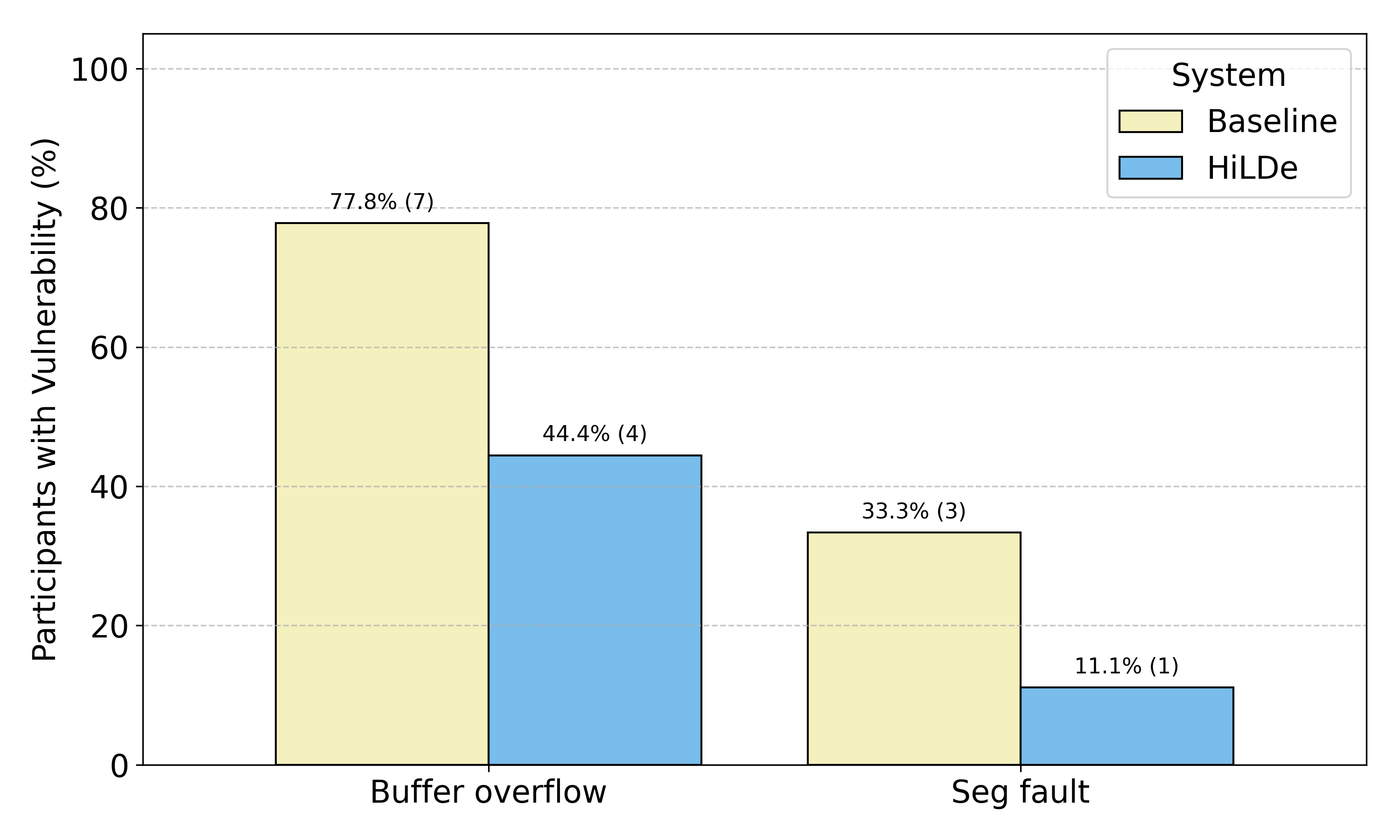}
        \caption{CSV File Write (T3) Mistakes}
        \label{fig:b_vulns}
    \end{subfigure}
    \hfill
    \begin{subfigure}{0.49\textwidth}
        \centering
        \includegraphics[width=\textwidth]{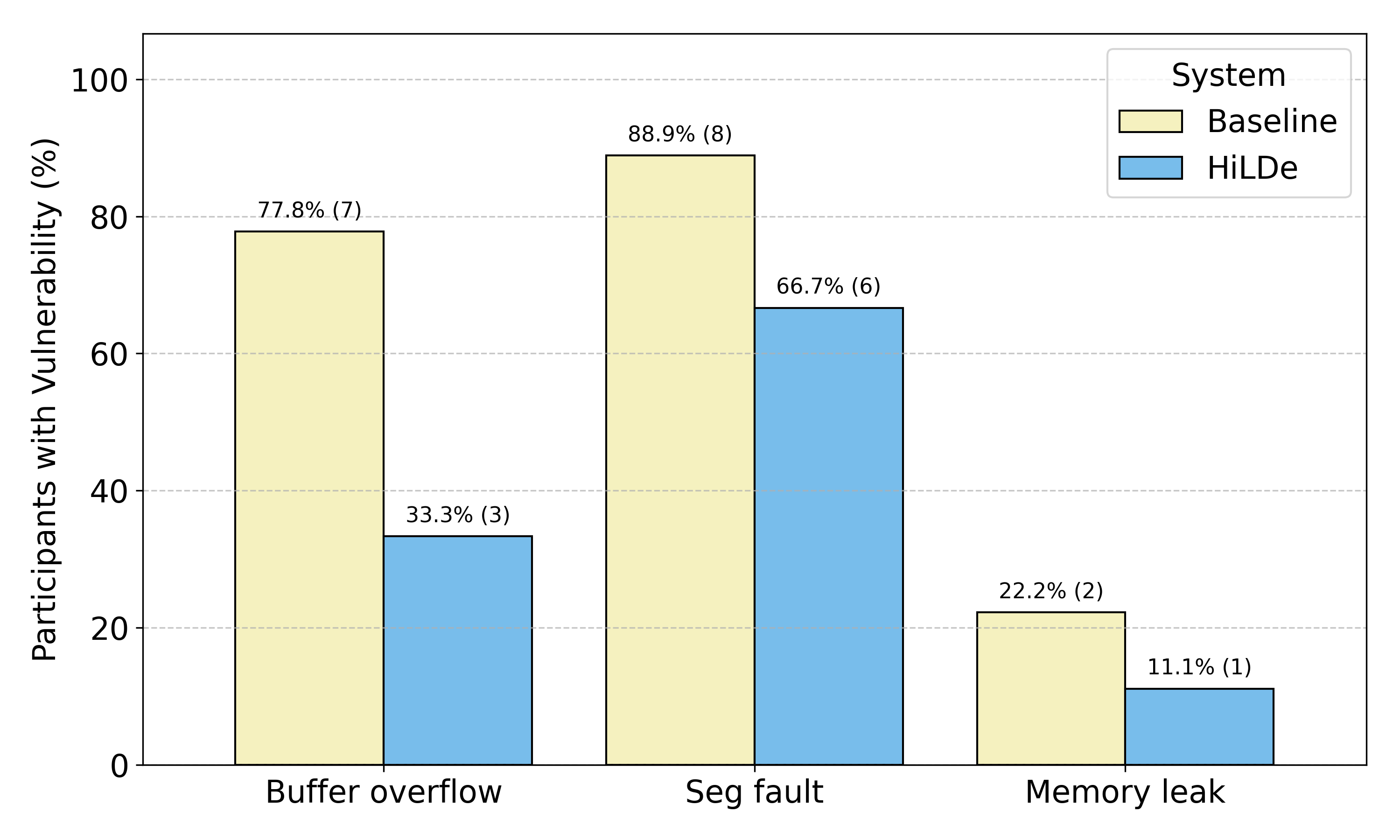}
        \caption{User Input to Struct (T4) Mistakes}
        \label{fig:d_vulns}
    \end{subfigure}
    \caption{Summary of all security vulnerabilities identified in \sys (blue) vs. \baseline (yellow) group solutions.}
    \label{fig:security_results}
\end{figure*}

%% file: secs/results.tex
\section{Results}
\label{sec:results}
In the following sections, we present a detailed quantitative and qualitative analysis using system log data and session transcripts corresponding our three research questions.
Our results highlight the effectiveness of \sys in helping programmers write secure code with LLMs, catch more mistakes in AI suggestions, and steer code generation to match their intent.

Participants successfully completed the task and passed all tests in 70 out of 72 instances.
While functional correctness was not our primary concern, we supplied a basic set of tests for each task to support participants and ensure some level of functionality.
These tests allowed participants to verify that their code compiled and view its output; all but two (P14: Task 2, P21: Task 1) were able to do so within 15 minutes.

We found no significant difference in cognitive load between \sys and \baseline, from responses to the NASA-TLX~\cite{hart1988development} metrics on a 5-point Likert scale.
Participants reported similar levels of mental demand ($M_\sys = 2.0$, $M_{\baseline} = 1.9$, $p=0.4$), temporal demand ($M_\sys = 1.9$, $M_{\baseline} = 1.6$, $p=0.1$), performance ($M_\sys = 4.3$, $M_{\baseline} = 4.4$, $p=0.8$), effort ($M_\sys = 2.1$, $M_{\baseline} = 2.0$, $p=0.2$), and frustration ($M_\sys = 1.7$, $M_{\baseline} = 1.6$, $p=0.2$).

\subsection{RQ1: Security}
\label{sec:results:rq1}
To investigate whether \sys helps participants write secure code, we compared the number of vulnerabilities present in the code written with each Assistant, as well as the number of intentional security repairs made.
Our overall results are shown in \autoref{fig:security_results} and \autoref{fig:repair_strategies}.

\subsubsection{\textbf{Participants wrote code with significantly fewer vulnerabilities using \sys}}
Participants using \sys generated code with 31\% fewer vulnerabilities on average compared to \baseline ($M_\sys = 2.67$,$M_{\baseline} = 3.89$,$p=0.01$, $r=0.53$).
These results are clearly seen in \autoref{fig:security_results}, where the number of vulnerabilities generated by participants using \sys (blue) is consistently lower than those using \baseline (yellow) across most tasks.
This difference is particularly impactful since participants reported having limited secure coding experience (average self-rating of 2.1 out of 5 for software security knowledge).

Across both Assistants, the most common mistakes were: insecure choice of \emph{hashing} algorithm (T1)%
\footnote{This vulnerability was illustrated in \autoref{sec:hilde-example}.},
unsafe source of \emph{randomness} from the \scode{ecdsa} library (T1), and \emph{symlink} vulnerabilities (T2).
Every solution written with \baseline used insecure password hashing practices.
Session logs showed that \baseline only suggested the \scode{SHA256} algorithm without salting, and participants did not discover resource-intensive alternatives like \scode{scrypt} or \scode{pbkdf2\_hmac} with salting, which were available in \sys.

In most cases, participants using \sys generated code with as many or fewer vulnerabilities than those using \baseline; the symlink vulnerability was the one exception:
although they often mitigated the risk of parent directory traversal by selecting the \scode{os.path.abspath} alternative to simply \scode{path.startswith}, none used \scode{os.path.realpath} to get the canonical path, leaving their code vulnerable to symlink attacks.
In one instance, P4 was able to correct the full path traversal vulnerability with \baseline, but only after consulting external documentation.

\subsubsection{\textbf{Participants using \sys intentionally corrected more vulnerabilities in AI-generated code}}
\label{sec:repair_strategies}
Due to LLM's nondeterministic nature, both Assistants occasionally generated code that was secure by default, requiring no participant intervention.
To better understand participant engagement, we distinguished between cases in which participants simply accepted already-secure code and cases where they \emph{intentionally} steered code generation towards a more secure alternative.

Out of 42 instances of intentional security repair, 71\% occurred with \sys while only 29\% occurred with \baseline(\autoref{fig:repair_strategies}).
Our analysis of session logs revealed two main strategies participants used to steer code generation: a UI-driven strategy, where participants selected a more secure alternative from the Assistant's suggestions (at the completion level for \baseline, and the token level for \sys); and a prompt-driven strategy, where participants explicitly prompted the Assistant to generate more secure code.
\sys participants were significantly more likely to use the UI-driven approach for security repairs, while \baseline participants relied more on explicit prompting ($percent_\sys=91$, $percent_\baseline=62$, $p=0.03$, $r=0.17$).

\input{figs/repair_strategies.tex}

\input{figs/time_distr.tex}
\subsection{RQ2: Overreliance}
\label{sec:results:rq2}

\subsubsection{\textbf{With \sys, participants spent more time evaluating LLM suggestions before accepting them}}
\label{sec:results:rq2:time}
In order to understand the level of critical thinking participants demonstrated with both Assistants,
we measured the average time each participant took to accept a completion after receiving it,
and the average time it took participants to finish a task.
We show the results in \autoref{fig:time_distr}.
Overall, we find that participants spend more time considering suggestions from \sys before accepting them compared to \baseline ($M_\sys = 76.20, M_{\baseline} = 35.55, t=-4.63, p<0.001$)
and they also take longer to submit their code when using \sys ($M_\sys = 514.98, M_{\baseline} = 396.39, t= -1.97, p=0.053$),
with the latter showing a trend toward significance but being marginally above the standard threshold.

Outside quantitative data, several participants (P5, P15, P18) mentioned that \sys encouraged them to more thoroughly evaluate the code they received from the LLM
and not take it at face value.
After seeing a number of hash functions suggested by \sys, P5 said that \emph{``if [they] had more time, [they] would Google each one to see which one is best''},
and wondered if \emph{``there was a safer way to do it''}.
P18 mentioned that ``\sys \emph{alternatives can be useful as long as you are not blindly tabbing and accepting everything''}.

\subsubsection{\textbf{\sys enabled users to understand the limitations of LLM-generated code}}

Participants expressed several ways in which the affordances and interaction model of \sys allowed them to understand its shortcomings,
and stopped them from blindly trusting the suggestions they get.

One common pitfall for \baseline participants (P3, P13, P16, P19, P21) was drawing an incorrect correlation between frequency of occurrence and correctness.
When seeing a particular pattern repeatedly in the global alternatives they thought \emph{``(P3) it is standard procedure''}, \emph{``(P16) is the default''} or
\emph{``(P21) is the only true way to do this''}.

This incorrect conclusion was less common among participants who used \sys.
A number of participants (P11, P16) seemed puzzled by the fact that the models' most likely solution was not the most correct, after seeing other alternatives.
P16 asked why the model \emph{``was not suggesting that in the first place. I am assuming most people use [unsafe alternative]\dots
I don't know what [safe alternative] does, I know what [unsafe alternative] does cause I know the basics \dots and maybe that applies to most other people,
which is what is feeding the AI''}.
This observation is a fair explanation for this behavior in language models.

\subsubsection{\textbf{After using \sys, participants had a more accurate sense of the correctness of their solutions}}

We found no significant difference in participants' self-reported confidence in the correctness of their solutions between Assistants ($M_\sys = 4.4$, $M_{\baseline} = 4.5$, $p=0.8$).
Participants did however report having significantly higher trust in AI generated code with \sys, though the effect size was small ($M_\sys = 3.39$, $M_{\baseline} = 3.72$, $p=0.02$, $r=0.23$).
Participants reported high levels of confidence in general,
and since those using \sys wrote code that was more secure,
their perceived correctness score was closer to their actual correctness score,
compared to those using \baseline.


Qualitatively, interactions of several participants (P15, P17, P21) with \sys indicate a better alignment between their real and perceived performance.
For instance, P15 performed the study in the \baseline-first setting, self-reporting a high level of confidence in their solutions after the first section,
while still accepting security vulnerabilities.
Then, after using \sys, they noted that it \emph{``[gave them] more alternatives''} that were \emph{``useful for recognizing security issues''}.
They recognized that \emph{``[they were] not familiar''} with the topic of the previous tasks and they \emph{``likely had security issues''},
also expressing that they wanted to \emph{``retroactively decrease [their] confidence score for} \baseline''.

\subsection{RQ3: Achieving Programming Goals with \sys}
\label{sec:results:programming_goals}

\emph{\textbf{With \sys, participants were able to achieve their programming goals more effectively}}
There were no significant differences between Assistants in self-reported control ($M_\sys = 3.8$, $M_{\baseline} = 3.8$, $p=0.5$), helpfulness of alternatives ($M_\sys = 3.9$, $M_{\baseline} = 4.0$, $p=0.6$), or understanding of alternatives ($M_\sys = 4.1$, $M_{\baseline} = 3.9$, $p=0.2$),
However, our qualitative analysis of instances from~\autoref{sec:repair_strategies} revealed that, at each LLM interaction, \sys guided participants through a three-step process:
\begin{enumerate}
    \item \textbf{Discover} alternative implementations that shape aspects of security, correctness, personal code style, etc.
    \item \textbf{Interpret} the alternatives using contextual explanations, requirements, and preferences to understand how each implementation aligns with their intent.
    \item \textbf{Act} on this understanding by selecting the most suitable implementation, effectively steering code generation.
\end{enumerate}
We illustrate this pattern through two in-depth case studies focusing on participants who had different types of intent.
Each case study details a participant's intentional interaction with the LLM, supplemented by similar experiences and a contrasting experience from the \baseline group.

\subsubsection{\textbf{Case Study 1---No explicit intent}}
P18 started solving Task 4 by requesting a completion from \sys, and immediately noticed uncertainty highlighting on the \scode{scanf} token: \emph{``I know with C, if you don't know what you're doing you can create some insecure stuff. So I'm assuming the LLM is just like asking if I want to use some more safe options."} (\textbf{Discover}).
P18 then opened the local alternatives for \scode{scanf}, reading the explanation for the first alternative which used the \scode{fgets} method: ``\scode{fgets} \emph{improves the safety by limiting the number of characters read and like yeah, buffer overflows and stuff."} (\textbf{Interpret}).
After consulting and ruling out the other---irrelevant and insecure---options, P18 selected the \scode{fgets} alternative: \emph{``Yeah, let's go with this."} and inspected how the completion changed (\textbf{Act}).

When asked to reflect on this interaction in the post-study interview, P18 said:
\emph{``It} [the model] \emph{wasn't sure of which function to use for user input and that's an important decision. It did make me aware of, yeah, we don't want to just accept arbitrary input."}
P11 had a similar experience: \emph{``I don't really think about the security of my code. But looking at the options, I was sort of motivated to pick something that would be more secure.''}
P16 echoed, ``\sys \emph{helped me better realize my intent instead of express my intent. It helped me realize my intent because I wasn't aware of it before.''}

After completing the same task with \baseline, P13 reflected: \emph{``I would say the only potential problem is I think I've heard that} \scode{scanf} \emph{is not secure or something, I don't know, maybe. But I guess for the purposes of this task, it didn't really matter."}
Although P13 reviewed \baseline's alternatives for reading user input—--including some secure options—--they ultimately accepted an insecure completion.

This response underscores how, without contextual affordances, P13 missed the opportunity to discover a safer alternative.
In contrast, P18, P11 and P16 also started out passive in their LLM interactions, but \sys encouraged them to discover and solidify their own programming goals.

\subsubsection{\textbf{Case Study 2---Well-defined intent}}
During the course of solving Task 2, P15 noticed that the test cases were failing because they
were printing an error message instead of raising an exception.
They went ahead and tried to prompt the model via a comment to \scode{\# raise exception instead of print}.
When they requested a completion from \sys at that location,
the model again suggested an unwanted print statement.
However, P15 noticed the highlighting in the first token of the line,
and \emph{``was curious if [they would] find''} an appropriate solution in the alternatives (\textbf{Discover}).
\sys in fact offered an exception clause as the first option.
They recognized this instantly (\textbf{Interpret}) and promptly accepted this alternative: \emph{``Great! here we go\dots''} (\textbf{Act}), and verified their solution passed the tests.

In the post-task interview they remarked, \emph{``having more alternatives show up was really nice,
because I could immediately see the raise exception thing, whereas previously I might have had
to re-prompt it a few times''}.
They found \sys particularly helpful since \emph{``they [knew] the issue, and }\sys\emph{was just showing how to solve it''}.
A number of participants (P3, P5, P13, P16) had similar observations,
assuring that they appreciated \emph{``(P3) [having] more granular control over code''}, 
being able to \emph{``(P16) make smaller changes''} and \emph{``(P5) line by line modifications''}.

On the other hand, participants had trouble accurately steering the \baseline to make fine-grained changes.
For instance, P14 found themselves in a similar situation as P15,
where they knew the correct function to use in their context, and prompted the \baseline for it through a comment.
The \baseline however \emph{``did not suggest at all''} their preferred implementation;
they had \emph{``difficulty getting} [\baseline]\emph{to use it''} and had to type it out themselves.
Other participants (P4, P11, P16, P19) also demonstrated this pattern of prompting via comments and not obtaining any acceptable suggestion.

Additionally, users (P3, P4, P5) complained about the inability to perform targeted changes with \baseline,
and only being able to \emph{``(P5) change the whole structure''}.
P3 compares both tools as follows: \emph{``In }[\baseline] \emph{if I am not satisfied with one of the middle steps,
I have to delete all the code below it and prompt [again]. But }\sys{}\emph{ \dots it will just generate code
based on the choice I made.''}.

These interactions clearly indicate the advantage of \sys when users engage with it with a clear goal in mind,
as \sys allows them to more precisely indicate their preferences at a granular level,
without the need to re-prompt or regenerate, like they would need to do with \baseline.
In the case of \sys, users have more agency in low-level LLM decision-making
and are able to align it with their personal goals.

\subsection{Recurrent \baseline Limitations}
\label{sec:results:programming_goals:baseline}

\baseline tended to generate alternatives that were not sufficiently diverse, and many participants (P5, P9, P14, P17, P21) found the alternatives \textbf{less helpful for discovering different implementations}.
During a post-task survey, P5 noted, \emph{``[Viewing alternatives somewhat helped because] the logic was the same, only in different format, like single line of multiple if-then-else, but the logic was the same.''}
P21 echoed this sentiment: \emph{``I don't know if} [the \baseline alternatives] \emph{helped that much. I picked the 1st one for almost all of them if they were available. I feel like there potentially could be alternates out there for sure."}

When more diverse alternatives were accessible, participants (P9, P16, P19, P21) found that \baseline was \textbf{less helpful for interpreting differences} between alternatives.
P19 was in the ``\sys-first'' group, and when they used \baseline, they immediately expressed a desire for \sys's local explanations:\emph{``I don't have...the comments that tell me what's the difference between the old one and the new one. So it makes it harder to understand,} continuing, \emph{``it takes me some time to actually see what has changed.''}

Once participants understood the differences, many (P9, P11, P13, P16, P17, P19, P21) felt that \baseline was \textbf{less helpful for deciding which alternative to accept}.
Participants often accepted insecure completions from more secure alternatives because they did not have enough context.
During a task P17 reflected, \emph{``the explanations would have at least given me idea. With} [\baseline{}] \emph{I was kind of left in the dark.''}
Even though P21 ultimately chose a safer alternative, they reflected: \emph{``I still don't know what \scode{safe\_load} means unless I go and open the docs. I didn't really know the trade-offs even though options were presented to me''.}

%% file: figs/repair_strategies.tex
\begin{figure}[h]
    \centering
    \includegraphics[width=0.49\textwidth]{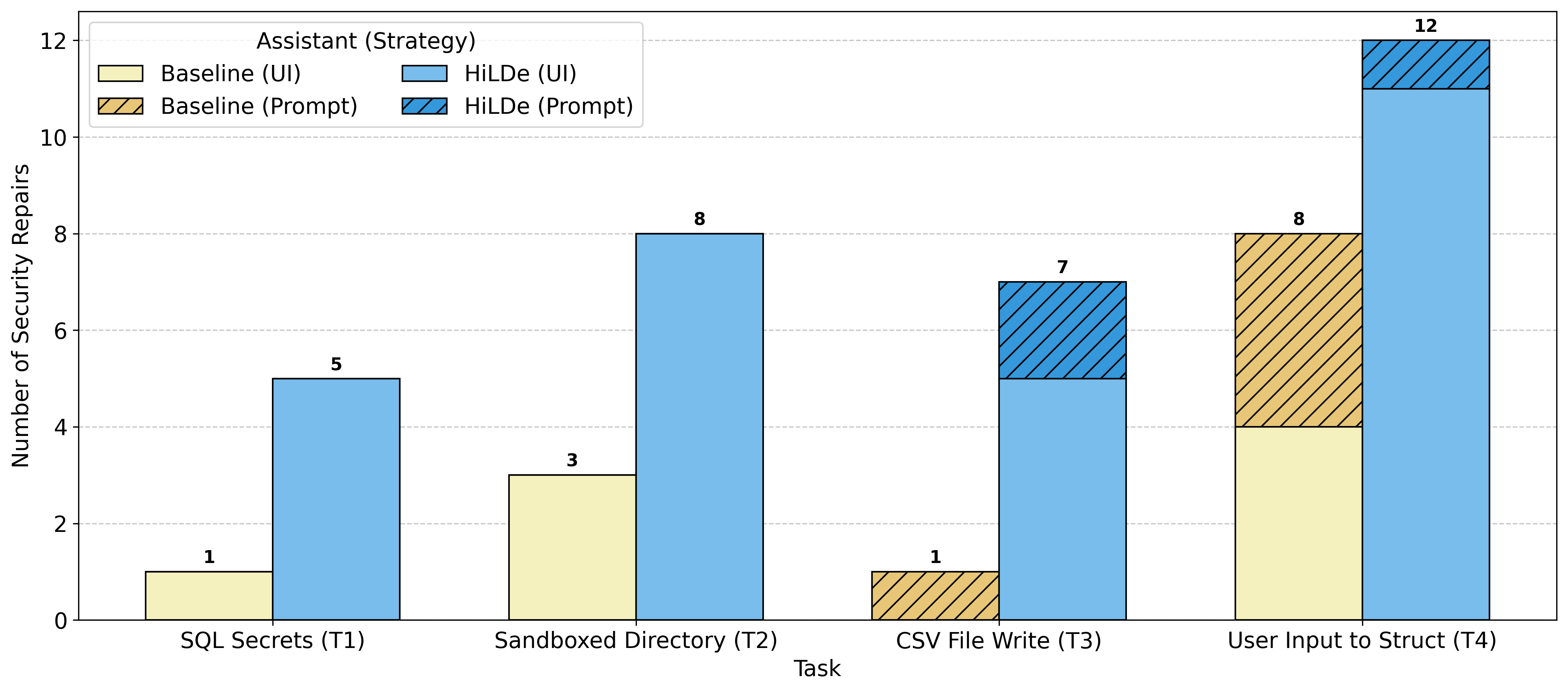}
    \caption{Intentional security repairs using \sys (blue) vs. \baseline (yellow) for each repair strategy for each task.}
    \label{fig:repair_strategies}
\end{figure}

%% file: figs/time_distr.tex
\begin{figure}[h]
    \centering
    \includegraphics[width=0.49\textwidth]{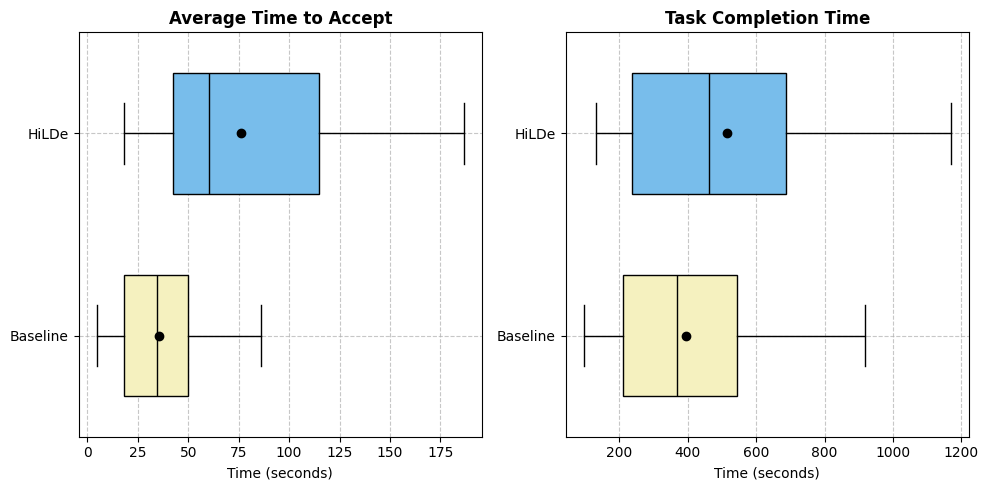}
    \caption{Distribution for average time to accept a completion (left) and task completion time (right) for \sys (blue) and \baseline (yellow)}
    \label{fig:time_distr}
\end{figure}

%% file: secs/discussion.tex
\section{Discussion and Future Work}
\subsection{The Benefit of \technique}
Our results show that \technique helps programmers explore and understand a rich space of implementation choices in LLM-generated code, unlike \baseline.

\autoref{sec:results:programming_goals:baseline} finds that participants became frustrated when \baseline generated many similar alternatives.
In standard decoding algorithms, models offer little variation when they are confident in their predictions.
Even if a wider variety of options were provided, it would be difficult to express every interesting choice within five global alternatives.
However, adding more global alternatives is impractical from a usability standpoint—--\autoref{sec:results:programming_goals:baseline} finds that participants already struggled to spot meaningful differences within the small set of \baseline alternatives, a challenge also reported by users of mainstream AI assistants~\cite{barke2023grounded}.

In contrast, \technique exposes less probable, but potentially valuable local alternatives, enabling users to discover a \emph{broad range} of options in a \emph{fine-grained} way.
This approach aligns with general calls for AI-resilient interfaces that help users understand the range of solutions an LLM can generate from a single prompt, forming more accurate mental models of LLM behavior~\cite{gu2024ai, gero2024supporting}.

\subsection{\technique Beyond Security}
We found that participants wrote significantly safer code using \sys, even though they were never explicitly instructed to prioritize security (\autoref{sec:results:rq1}).
However, \sys also helped participants achieve programming goals beyond security.

For example, during Task 1, P21 noticed a SQL server \scode{cursor} declaration highlighted in red and opened the local alternatives view. 
After seeing the second alternative, they declared, \emph{``Oh yeah, usually I actually use this...the more idiomatic way is to use \scode{with} so that it automatically closes for you.''} 
P21 selected this alternative, confirmed that the downstream output matched their intuition, and then continued solving the task. 
In this case, \sys helped P21 recognize their personal coding practices in the alternatives and steer the completion to match their preferences.

While the current implementation targets security, this approach can be easily adapted for other priorities such as efficiency, maintainability, energy conservation, or personal coding style;
The underlying prompt for \sys is slightly tuned to surface security-related decision points, but adjusting the prompt could make \sys more general or target other code attributes. 
For example, P9 saw the utility of \sys in different contexts, \emph{``whether you want to make your code secure, whether you want to make your code readable''}.
We are excited to explore a more general, customizable version of \sys in future work.

\subsection{Intentionality vs. Efficiency in Human-AI Collaboration}
Our study highlights a key trade-off: while \sys slowed down task completion (\autoref{sec:results:rq2:time}), this additional time facilitated more intentional LLM code review and decision-making—--which was often at odds with participants' usual preference for speed. 
For example, P13 appreciated \sys for complicated tasks where they have to ``\emph{make multiple choices}", yet found it ``\emph{more cumbersome}" for simple code. 
Others (P3, P5, P16, P17, P19, P21) saw the utility in \sys for unfamiliar domains, but preferred \baseline when they ``\emph{(P5) want answers fast}".
This distinction parallels the ``exploration" and ``acceleration" modes that Barke et al.~\cite{barke2023grounded} identified among users of AI programming assistants.

In prior user studies, task speed has often served as a measure of AI assistant utility~\cite{peng2023impact, kuttal2021trade, ziegler2024measuring, pu2025assistance}, but our findings highlight how interface design can encourage more intentional code generation and help programmers strike a balance between speed and code quality.

\subsection{\sys Limitations}
During study sessions, \sys sometimes made unanticipated changes to downstream code after participants selected a local alternative.
In a post-task survey, P14 reflected \emph{``It was helpful to be able to change options, although, when it regenerated it just removed some of the things.''}
Most participants re-applied their previous selections, but some found this to be too much work, choosing to leave the completion as is.

The problem is that when a user selects a local alternative, \sys prompts the underlying model to generate a new completion from that point, which can overwrite prior downstream edits.
To mitigate this (c.f. \autoref{sec:hilde-implementation}) \sys requests several completions from the model and chooses the one most similar to the original code.
However, if the user selects a token with very low-likelihood of being chosen by the model, it is unlikely that any of the generated completions will exactly match their previous edit, resulting in that change being lost.
Ensuring that AI assistants preserve incremental user modifications remains an open challenge for future work~\cite{kazemitabaar2024improving}.

Additionally, a few participants (P7, P11, P18) found \sys's interaction method and token highlighting overwhelming at times, especially when it surfaced local alternatives that were not useful to them. 
As P7 remarked, \emph{``That amount of cognitive overhead ended up...hindering my own thought process...because there were more possibilities to consider that I didn't need to consider''}. 
However, we found no statistically significant difference in cognitive load between the two assistants.